\documentclass[twocolumn,english,prl,aps,showpacs,superscriptaddress,nofootinbib,longbibliography,floatfix]{revtex4-1}

\usepackage[english]{babel}

\usepackage{verbatim}
\usepackage{amsmath}
\usepackage{amssymb}
\usepackage{graphicx}
\usepackage{esint}
\usepackage{soul}

\usepackage{comment}

\makeatletter

\usepackage{bm}
\usepackage{xcolor}
\usepackage{subfigure}
\usepackage{array}

\@ifundefined{textcolor}{}{%
 \definecolor{BLACK}{gray}{0}
 \definecolor{WHITE}{gray}{1}
 \definecolor{RED}{rgb}{1,0,0}
 \definecolor{GREEN}{rgb}{0,1,0}
 \definecolor{BLUE}{rgb}{0,0,1}
 \definecolor{CYAN}{cmyk}{1,0,0,0}
 \definecolor{MAGENTA}{cmyk}{0,1,0,0}
 \definecolor{YELLOW}{cmyk}{0,0,1,0}
}

\usepackage{bm}
\usepackage{float}

\newcommand{\be}{\begin{equation}}
\newcommand{\ee}{\end{equation}}
\newcommand{\bea}{\begin{eqnarray}}
\newcommand{\eea}{\end{eqnarray}}
\newcommand{\beq}{\begin{equation}}
\newcommand{\eeq}{\end{equation}}
\def\lb{\label}

\newcommand{\Ch}{\chi^{(3)}}

\newcommand{ \Po}{P^{(3)}}

\newcolumntype{x}[1]{>{\centering\arraybackslash}p{#1}}

\makeatother

\def\nn{\nonumber}
\def\lb{\label}
\def\pref#1{(\ref{#1})}

\newcount\bozza \bozza=0
\ifnum\bozza=1
\newdimen\shift \shift=-2truecm
\def\lb#1{%
{\label{#1}\rlap{\kern\shift{$\scriptstyle#1$}}}}
\else\def\lb#1{\label{#1}} \fi

\def\o{\omega}

\def\O{\Omega}

\begin{document}

\title{Terahertz ionic Kerr effect: Two-phonon contribution to the nonlinear
optical response in insulators}

\author{M. Basini}
\affiliation{These authors contributed equally to this work}
\affiliation{Department of Physics, Stockholm University, 10691 Stockholm, Sweden}
%\affiliation{These authors contributed equally to this work}

\author{M. Udina}
\affiliation{These authors contributed equally to this work}
\email{mattia.udina@uniroma1.it}
\affiliation{Department of Physics and ISC-CNR, ``Sapienza'' University of Rome, P.le
A. Moro 5, 00185 Rome, Italy}
%\affiliation{These authors contributed equally to this work}

\author{M. Pancaldi}
\affiliation{Department of Molecular Sciences and Nanosystems, Ca' Foscari University of Venice, 30172 Venice, Italy}
\affiliation{Elettra-Sincrotrone Trieste S.C.p.A., 34149 Basovizza, Trieste, Italy}

\author{V. Unikandanunni}
\affiliation{Department of Physics, Stockholm University, 10691 Stockholm, Sweden}

\author{S. Bonetti}
\affiliation{Department of Physics, Stockholm University, 10691 Stockholm, Sweden}
\affiliation{Department of Molecular Sciences and Nanosystems, Ca' Foscari University of Venice, 30172 Venice, Italy}

\author{L. Benfatto}
%\email{lara.benfatto@roma1.infn.it}
\affiliation{Department of Physics and ISC-CNR, ``Sapienza'' University of Rome, P.le
A. Moro 5, 00185 Rome, Italy}
\date{\today}

\begin{abstract}
The THz Kerr effect measures the birefringence induced in an otherwise isotropic material by a strong THz pulse driving the Raman-active excitations of the systems. Here we provide experimental evidence of  
a sizable Kerr response in insulating SrTiO$_3$ due to infrared-active lattice vibrations. Such a signal, named ionic Kerr effect, is associated with the simultaneous excitation of multiple phonons. Thanks to a theoretical modeling of the time, polarization and temperature dependence of the birefringence we can disentangle the ionic Kerr effect from the off-resonant electronic excitations, providing an alternative tunable mechanism to modulate the refractive index on ultrashort time-scales via infra-red active phonons.
\end{abstract}

\maketitle

The latest advances in the generation of intense terahertz (THz) field pulses made it possible to investigate the low-frequency counterpart of nonlinear optical phenomena in condensed matter, conventionally studied with visible light, as it is the case for the THz Kerr effect \cite{sajadi15,hoffmann09terahertz,melnikov22terahertz}. The DC Kerr effect detects a birefringence in an otherwise isotropic material proportional to the square of the applied DC electric field, and it is a standard  measurement of the {third-order $\chi^{(3)}$ non-linear optical response of the medium \cite{boyd_book}. Basically, the four-wave mixing between the AC probe $E_{AC}(\omega)$  and the DC pump $E_{DC}$ field leads to a non-linear polarization $P^{(3)}\sim \chi^{(3)} E_{DC}^2 E_{AC}$ (space indexes are omitted). The $P^{(3)}$ in turn modulates the refractive index at the same frequency $\omega$ of the AC field, with a space anisotropy set by the direction of $E_{DC}$. In its optical counterpart, the spectral components around zero frequency of the squared AC field play the same role of the DC component. More recently, THz and optical pulses have been combined in a pump-probe setup
%a \textcolor{blue}{four-wave mixing (RIPETIZ)} among THz and optical pulses has been exploited 
to measure the so-called THz Kerr effect \cite{hoffmann09terahertz}. 
The main advantage over  its all-optical counterpart is that intense THz pump pulses can strongly enhance the signal  
by matching Raman-like low-lying excitations in the same frequency range, such as lattice vibrations \cite{kampfrath_prl17,johnson_prl19,korpa18mapping, frenzel23}, or collective-modes in broken-symmetry states, as for magnetic \cite{kampfrath11coherent,baierl16nonlinear,bonetti16thzdriven,mashkovich19terahertz,mashkovich21terahertz} or superconducting transitions \cite{shimano_prl18,shimano_prb20,gallais_npj22}. Such a resonant response usually adds up to the background response of electrons, and it can be used to identify the microscopic mechanisms underlying the coupling among different degrees of freedom. 
\begin{figure}[t]
\centering
\includegraphics[width=8.5cm
, clip=true]{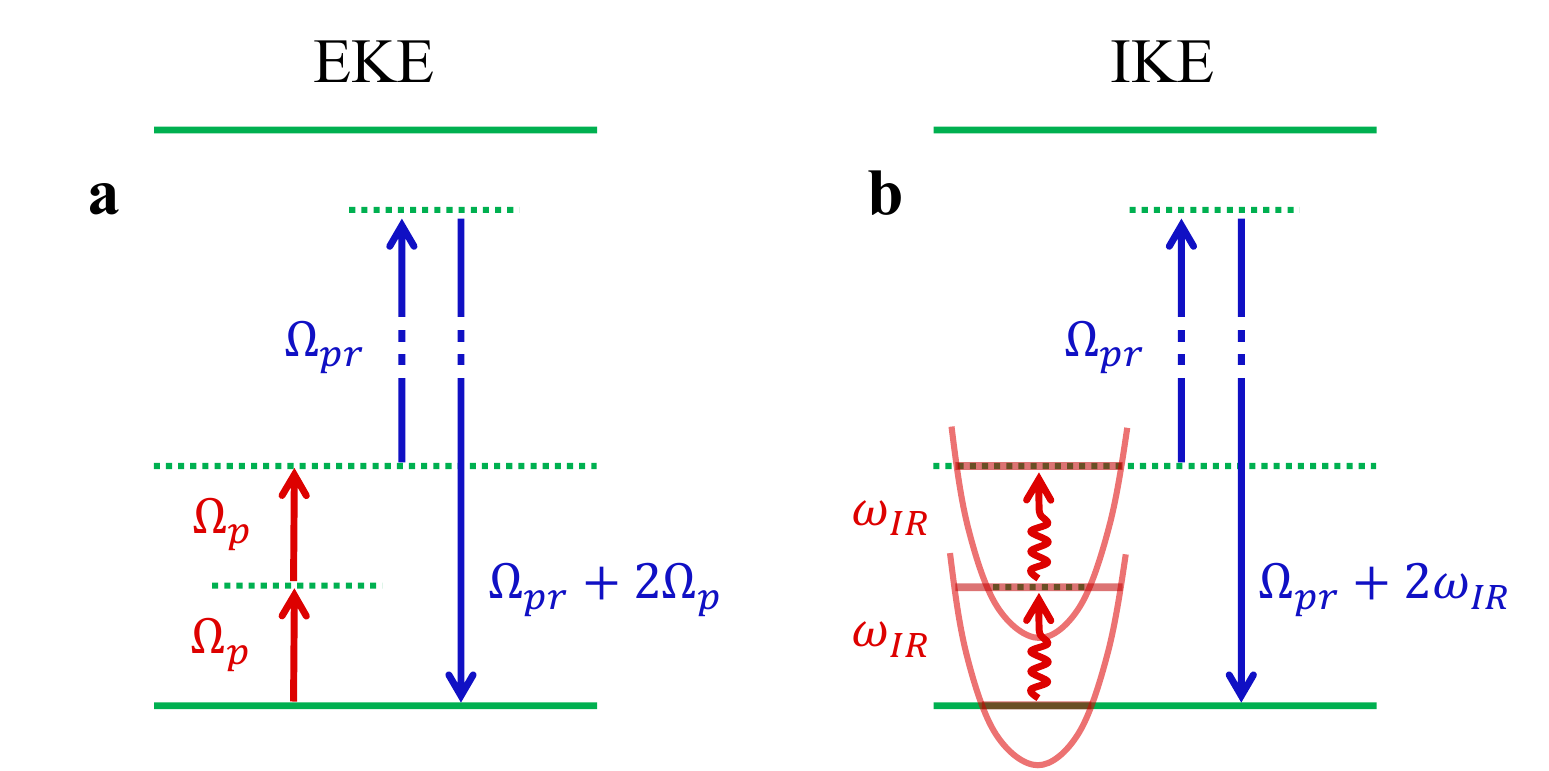}
\caption{Terahertz electronic and ionic Kerr effect in a wide-band insulator. In the THz Kerr effect the pump field (red arrows) drives an intermediate state by a sum-frequency two-photons  process. Such state subsequently scatters the visible probe field (blue arrows). In the EKE (\textbf{a}) the intermediate electronic state (dashed line) is virtual, being a insulator, so it relaxes back almost instantaneously giving a $2\Omega_p$ modulation of the emitted light. In the IKE (\textbf{b}) two IR-active phonons (wavy lines) can be used to reach the virtual electronic state (process labeled as (a) in the text), leading to a modulation at $2\o_{IR}$. This pathway only occurs when the THz pulse is resonantly tuned to the phonon frequency, i.e.\ when $\o_{IR}\simeq \O_p$.} 
\label{intro}     
\end{figure}

As a general rule, the THz Kerr response, scaling as the THz electric field squared, is not affected by infrared-active (IR-active) phonons, that correspond to lattice displacements that are linear in the applied electric field. While this is strictly true  to first order \cite{nelson_sci},  higher-order processes are not excluded by symmetry.
%even though no experimental evidence for such an alternative contribution to the THz Kerr effect has been provided so far. 
 In this work, we demonstrate that the non-linear excitation of the IR-active soft phonon mode in the archetypal perovskite SrTiO$_3$ (STO) leads to a sizeable contribution to the Kerr signal,  which we name ionic Kerr effect (IKE). 
STO is a quantum paraelectric \cite{saxena_STO_natphys14} showing a large dielectric constant at room temperature ($\epsilon_0\approx 300$), and with a low-lying phonon mode that progressively soften as the temperature is lowered \cite{bonetti_natphys,marsik16terahertz,vogt95}. Such excitation is an IR-active transverse optical (TO$_1$) phonon, and it is the same phonon mode which is responsible for the paraelectric to ferroelectric transition upon, e.g., Ca doping \cite{benhia_STO_natphys17}. Furthermore, it has been recently pointed out its possible role in the superconducting transition in electron-doped samples \cite{balatsky_prl15, gastiasoro_review_STO_annphys22, gastiasoro_cm22}, and the possibility of realizing dynamical multiferrocity upon driving it with circularly polarized THz electric fields \cite{basini_DM, DM_theory}. 
Here we show that  besides the well-studied electronic Kerr effect (EKE), due to off-resonant electronic transitions in wide-band insulating STO \cite{stryland} (Fig.\ \ref{intro}\textbf{a}), a ionic contribution,  associated with the second-order excitation of the TO$_1$ phonon, is present (Fig.\ \ref{intro}\textbf{b}). Such IKE  manifests itself with a sizable temperature dependence of the Kerr response, which is unexpected for the EKE in a wide-band insulator. In contrast, the IKE rapidly disappears by decreasing temperature, due to the phonon frequency softening far below the central frequency of the pump field. We are able to clearly distinguish between EKE and IKE thanks to a detailed theoretical description of the THz Kerr signal, which we retrieve experimentally as a function of the light polarization and of the pump-probe time delay $t_{pp}$. So far, the  regime of large lattice displacements has been mainly investigated to exploit the ability of infrared-active vibrational modes, driven by strong THz pulses, to anharmonically couple to other phononic excitations \cite{maehrlein_prb18,vonhoegen18probing,forst11nonlinear,subedi14theory,nelson_sci}.  Our work demonstrates that non-linear phononic represents also a suitable knob to manipulate the refractive index of the material,  providing potentially  an additional pathway to drive materials towards metastable states which may not be accessible at thermal equilibrium \cite{cavalleri_review_natphys22}.

\section{Experimental setup}
Measurements are performed on a $500\ \mu$m-thick SrTiO$_3$ crystal substrate (MTI Corporation), cut with the [001] crystallographic direction out of plane. Broadband single-cycle THz radiation is generated in a DSTMS 
%(dimethylamino-methyl-stilbazolium trimethylbenzenesulfonate) 
crystal  via optical rectification of a 40 fs-long, 800 $\mu$J near-infrared laser pulse centered at a wavelength of 1300 nm. The near-infrared pulse is obtained by optical parametric amplification from a 40 fs-long, 6.3 mJ pulse at 800 nm wavelength, produced by a 1 kHz regenerative amplifier. As schematically shown in Fig.\ \ref{setup}, the broadband THz pulses are filtered with a 3 THz band-pass filter, resulting in a peak frequency of $\O_p/2\pi=3$ THz (with peak amplitude of 330 kV/cm),
%with half width at half maximum of $\sim 0.65$ THz, 
and focused onto the sample to a spot of approximately 500 $\mu$m in diameter.
%The narrow band terahertz radiation is obtained by filtering the terahertz field with a 3 THz band-pass filter, resulting in a peak frequency of $\O_p/2\pi=3$ THz and a full width at half maximum of 0.5 THz \textcolor{red}{(dalla FT del segnale in tempo a me viene piu' largo)}. 
The time-delayed probe beam, whose polarization is controlled by means of a nanoparticle linear film polarizer, is a 40 fs-long pulse at 800 nm wavelength normally incident onto the sample surface. A 100 $\mu$m-thick BBO crystal ($\beta$-BaB$_2$O$_4$, Newlight Photonics) and a shortpass filter are used for converting the probe wavelength to 400 nm and increasing the signal-to-noise ratio in temperature-dependent measurements. The probe size at the sample is approximately 100 $\mu$m, substantially smaller than the THz pump. The half-wave plate (HWP) located after the sample is used to detect the polarization rotation of the incoming field induced by the non-linear response, and a Wollaston prism is used to implement a balanced detection scheme with two photodiodes. The signals from the photodiodes are fed to a lock-in amplifier, whose reference frequency (500 Hz) comes from a chopper mounted in the pump path before the DSTMS crystal.
\begin{figure}[t]
  \includegraphics[width=8.5cm,clip=true]{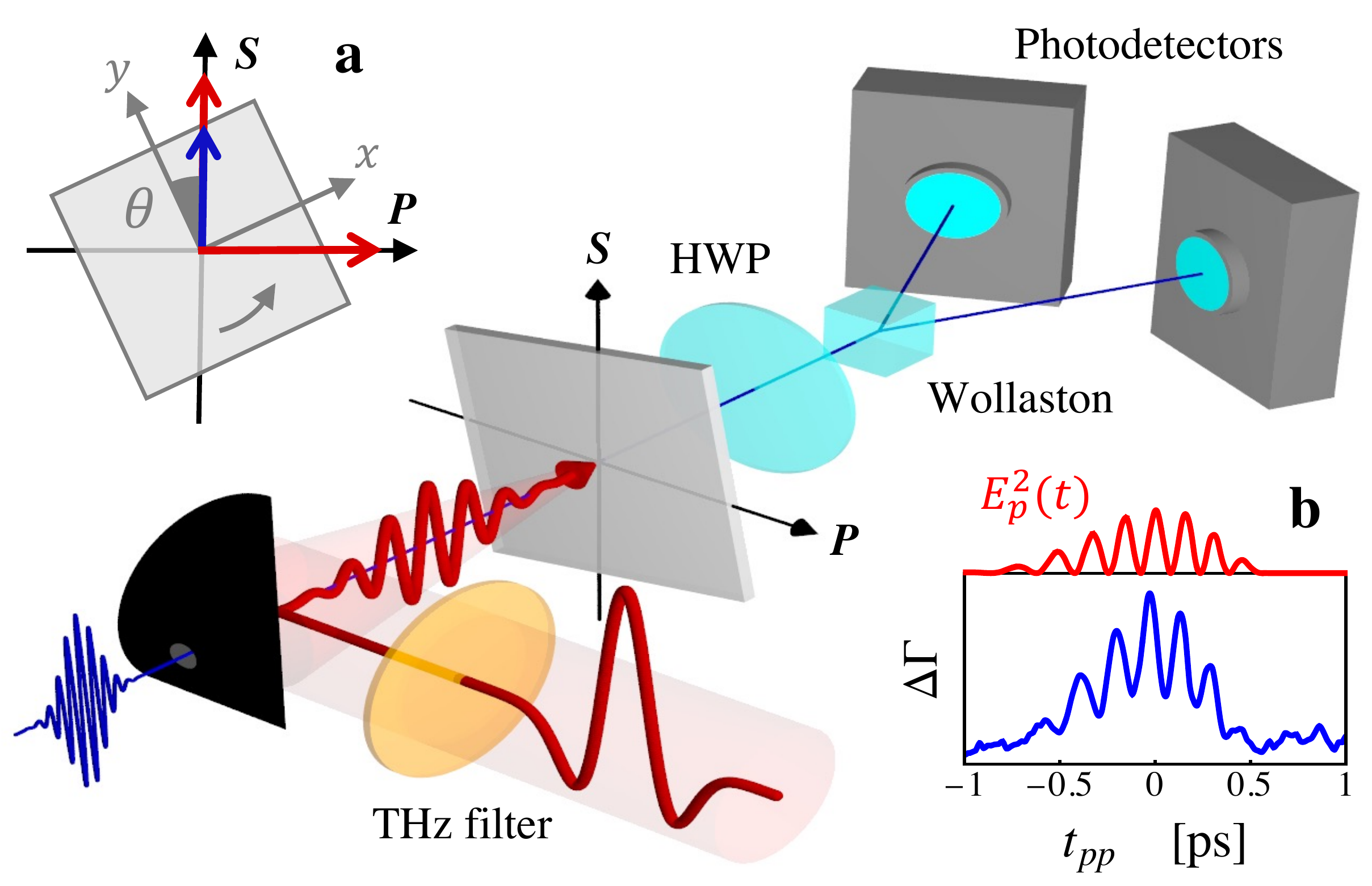}
\caption{Schematics of the experimental setup. The THz pulse is in red while the optical probe is in blue. Inset \textbf{a}: schematic of the polarization geometry. The sample is rotated in the $(S,P)$ plane, such that the $y$ crystallographic direction forms a variable angle $\theta$ with respect to $S$. Here the probe (blue arrow) polarization is fixed along $S$, while the pump pulse (red arrow) is polarized along either $S$ or $P$ for linearly polarized light and in both directions for circularly polarized light.  Inset \textbf{b}: typical time-trace of $\Delta\Gamma$ at fixed $\theta = 67.5^{\circ}$ (blue line) compared with the intensity profile of the linear pump pulse (red line). }    
\label{setup}                                               
\end{figure}  
Data are collected as a function of the $t_{pp}$ time-delay as well as a function of the angle $\theta$ that the probe polarization direction forms with respect to the main crystallographic axes. For convenience, here we assume that the probe beam is kept fixed along the $S$ direction, such that $\mathbf{E}_{pr}(t) = \left( E_{pr}(t)\sin{\theta}, E_{pr}(t)\cos{\theta} \right)$, while the pump field $\mathbf{E}_p$ can be either linearly polarized (along $S$ or $P$ direction) or circularly polarized \cite{basini_DM}, by placing a quarter-wave plate after the band-pass filter (see Appendix \ref{appA}), such that in general
\be
\lb{pump}
\mathbf{E}_p(t) = \begin{pmatrix}E_{p,S}(t) \sin{\theta}+E_{p,P}(t)\cos{\theta}\\E_{p,S}(t) \cos{\theta}-E_{p,P}(t)\sin{\theta} \end{pmatrix}.
\ee
In the presence of Kerr rotation, the probe field transmitted through the sample $\mathbf{\tilde E}(t)$ acquires a finite $P$ component. The half-wave plate rotates $\mathbf{\tilde E}(t)$ by $45^{\circ}$ with respect to $P$ direction and the outgoing signal reaching the two photodetectors $(\Gamma_1, \Gamma_2)$  reads \cite{jimenez}
%The behavior of the waveplate can be modeled as a linear combination of the outgoing signal $(\Gamma_1, \Gamma_2)$ along the main axes of the waveplate as a function of the sample polarization along $(S, P)$, i.e.\
%
\be
\lb{waveplate}
\begin{pmatrix} \Gamma_1 \\ \Gamma_2 \end{pmatrix} \propto \begin{pmatrix}1 & 1 \\ 1 & -1 \end{pmatrix}  \begin{pmatrix} \tilde E_S \\ \tilde E_P \end{pmatrix}.
\ee
In a pump-probe detection scheme, the use of a chopper on the pump path allows one for measuring, with a lock-in amplifier, the differential intensity $|\Gamma_1|^2- |\Gamma_2|^2$ with (\emph{on}) and without (\emph{off}) the pump, to obtain 
\be
\Delta\Gamma = \Delta\Gamma_{\textrm{\emph{on}}} -\Delta\Gamma_{\textrm{\emph{off}}} \propto (\tilde E_S \tilde E_P)_{\textrm{\emph{on}}}- (\tilde E_S \tilde E_P)_{\textrm{\emph{off}}}.
\lb{Dgamma}
\ee
A typical time-trace of $\Delta \Gamma$ for a linearly polarized pump pulse along $P$ and at fixed angle $\theta$ is shown in Fig.\ \ref{setup}\textbf{b}, together with $E^2_p(t)$. As we shall see below, the close qualitative correspondence among the two signals is a direct consequence of pumping a band insulator below the band gap. To measure the angular dependence of the response, we then choose $t_{pp}$ at the maximum $\Delta \Gamma$ amplitude, and we record its change as a function of $\theta$. The results are shown in Fig.\ \ref{polariz}\textbf{a} for both linearly polarized, in either the parallel and the cross-polarized configuration, or circularly polarized pump pulses. In all cases, the signal displays a marked four-fold angular dependency, along with a smaller two-fold periodicity. 
\begin{figure}[t]
\centering
\includegraphics[width=8.5cm, clip=true]{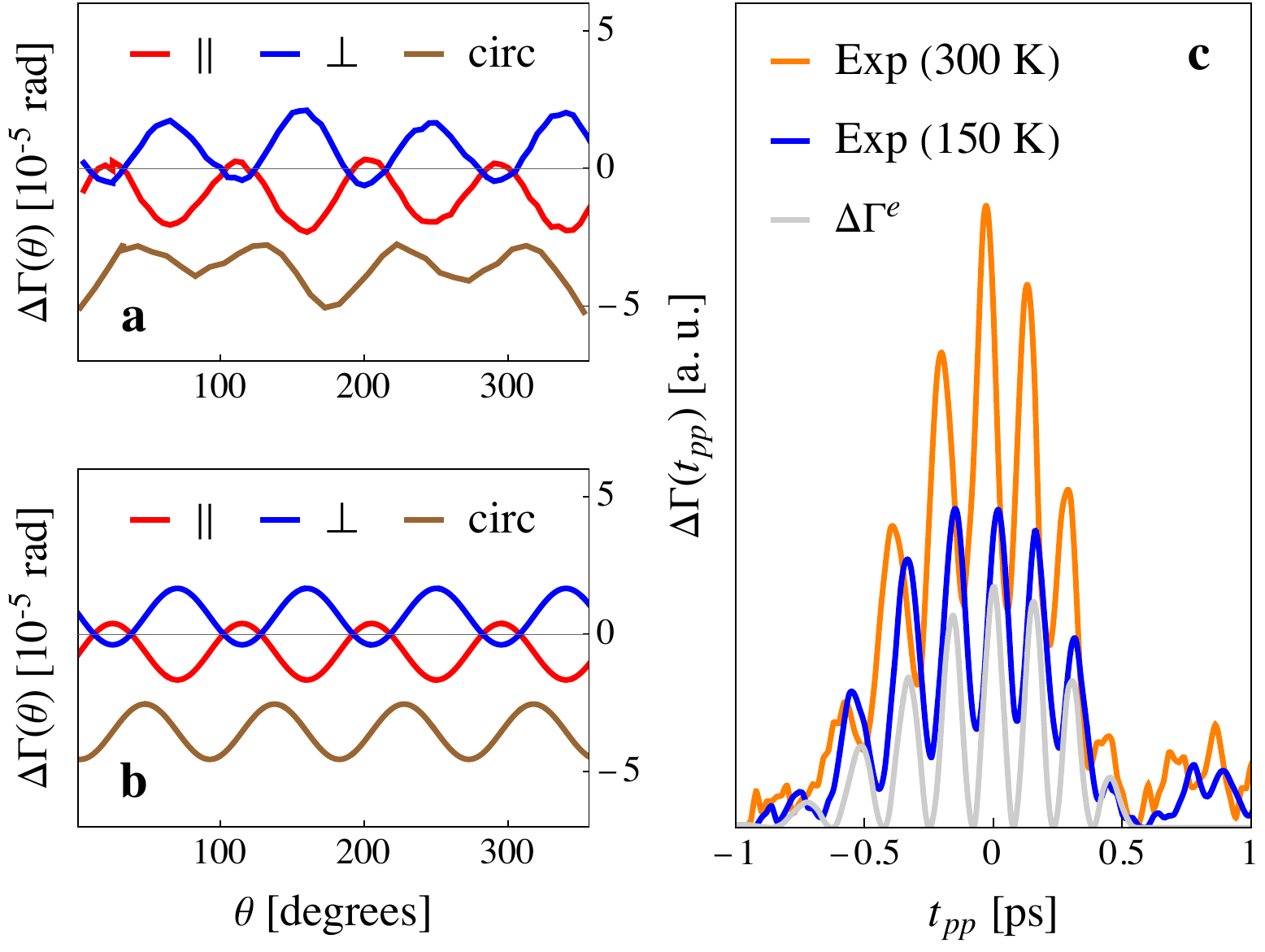}
\caption{Polarization and time dependence of the THz Kerr effect. \textbf{a} Experimental data as a function of $\theta$ at 300 K and $t_{pp}\simeq 0$, where the signal is maximized, in the parallel ($\parallel$, red curve), cross-polarized ($\perp$, blue curve) and circular (circ, brown curve) configuration. \textbf{b} Corresponding simulations from Eq.\ \pref{finalEL}, accounting for a finite polarization misalignment $\Delta_\theta=5^{\circ}$ between the pump and probe pulses. \textbf{c} Experimental data as a function of $t_{pp}$ at fixed angle $\theta = 67.5^{\circ}$ in the $\perp$ configuration at $300$ K (orange curve) and 150 K (blue curve), compared with the simulated off-resonant electronic contribution $\Delta\Gamma^e$ (gray curve).}  
\label{polariz}                         
\end{figure}  

\section{Electronic Kerr effect}
In order to assess such angular dependence, we first outline the theoretical description of the conventional EKE.  The transmitted probe field $\mathbf{\tilde E}(t)$ will contain both linear and non-linear components with respect to the total applied field $\mathbf{E}_p+\mathbf{E}_{pr}$. Since the detection is restricted around the frequency range of the visible probe field, the linear response to the pump can be discarded, as well as higher-harmonics of the probe generated inside the sample. We can then retain for $\tilde E_S \sim P^{(1)}_S$, with $P^{(1)}_S$ the linear response to the probe along $S$, while $\tilde E_P$ scales as the third-order polarization for a centro-symmetric system, i.e.\ $\tilde E_P \sim P^{(3)}_P$. By decomposing $P^{(3)}_P$ in its $(x, y)$ components one gets:
\be
\Delta\Gamma \propto \tilde E_S P_P^{(3)} = \tilde E_S \left [ \cos \theta P^{(3)}_x - \sin \theta P^{(3)}_y \right].
\lb{Dgamma2}
\ee
As mentioned above only contributions linear in $E_{pr}$ to $P^{(3)}_P \simeq \chi^{(3)}E_{pr} E_{p}^2$ must be retained, where $\chi^{(3)}$ is the third-order susceptibility tensor. By making explicit the space and time dependence one can finally write:
%, which can be seen as an induced sideband contribution to the spectrum of the optical pulse, oscillating around $\O_{pr}\pm2\O_{p}$ and generated by the interaction with the intense THz wave. 
%Note in passing that here we are directly working with the e.m.\ fields rather then with the related vector potentials for the sake of an explicit comparison between data and simulation, even if the two descriptions are fully equivalent from a theoretical perspective, as shown in Section \ref{sec2.1}. 
%
\be
\begin{aligned}
\lb{polar}
\Po_i(t, t_{pp}) = &\int dt' \sum_{jkl} E_{pr,j}(t) \times \\ \times & \chi^{(3)}_{ijkl}(t+t_{pp}-t') \bar E_{p,k}(t') \bar E_{p,l}(t'),
\end{aligned}
\ee
where we introduced the time shift $t_{pp}$ between the pump and probe pulses, such that $\bar E_{p}(t+t_{pp}) \equiv E_{p}(t)$,  with $\bar E_{p}(t)$ and $E_{pr}(t)$ centered around $t=0$ (see Refs.\ \cite{udina_prb19, huber21ultrafast} for a similar approach). Since the pump frequency is much smaller than the threshold for electronic absorption ($E_g \sim3$ eV), the nonlinear electronic response should be ascribed to off-resonant interband excitations, leading to a nearly instantaneous contribution to the third-order susceptibility tensor $\chi^{(3)}$, which can be well approximated by a Dirac delta-function, i.e.\ $\Ch_{ijkl}(t) \sim \chi^{(3)}_{ijkl}\delta(t)$, with $\chi^{(3)}_{ijkl}$ a constant value. Under this assumption, Kleinman symmetry for the cubic $m3m$ class appropriate for STO above $T_s\simeq105$ K assures that the off-diagonal components of the third-order susceptibility tensor have the same magnitude \cite{boyd_book}. In the specific case of insulating SrTiO$_3$, at room temperature one finds $\Ch_{iijj} = \Ch_{ijji} = \Ch_{ijij} \simeq 0.47 \Ch_{iiii}$ \cite{weber_book}. Finally, since the measurement at the photodetectors corresponds to the time-average field intensity $|\Gamma_i|^2$, one can replace $E_{pr,j}(t)$  in Eq.\ \pref{polar} with its value at $t=0$, so that the probe pulse acts as an overall prefactor, relevant only for its polarization dependence. With straightforward algebra, and using the decomposition \pref{pump} and \pref{Dgamma2}, one obtains the general expression for the EKE:
\bea
\lb{finalEL}
\Delta\Gamma^e(t_{pp},\theta) \propto \frac{1}{4} \Big[  \bar E_{p,P}^2(t_{pp}) - \bar E_{p,S}^2(t_{pp}) \Big] \Delta\chi \sin{(4\theta)}+\nn \\
+  2 \bar E_{p,P}(t_{pp})\bar E_{p,S}(t_{pp}) \Big[ \Ch_{xxyy}+ \frac{1}{2}\Delta\chi \sin^2{(2\theta)} \Big],
\eea
where $\Delta\chi \equiv \Ch_{xxxx}-3\Ch_{xxyy}$. Eq.\ \pref{finalEL} accounts very well for the angular dependence of the signal reported in Fig.\ \ref{polariz}\textbf{a}. 
In particular, for linearly polarized pump pulses, one recovers in Fig.\ \ref{polariz}\textbf{b} the four-fold symmetric modulation, and the overall sign change observed when going from the parallel ($\bar E_{p,P}=0$) to the cross-polarized configuration ($\bar E_{p,S}=0$). 
Within the same formalism, a finite polarization misalignment $\Delta_{\theta}$ between the pump and probe pulses with respect to $S
$ has been considered to explain the small overall (positive or negative) shift of the signal around which the oscillations occur (see Appendix \ref{appB}).
When instead circularly polarized pump pulses are applied, the mixed term $\bar E_{p,P}\bar E_{p,S}$ in Eq.\ (\ref{finalEL}) is different from zero and carries a finite isotropic contribution leading to a sizable vertical shift of the signal (even if $\Delta_{\theta}=0$), which has been also reported in Fig.\ \ref{polariz}\textbf{a}. Notice that the difference signal between right-handed and left-handed circularly polarized light, as the one measured in Ref.\ \cite{basini_DM}, picks up only the $E_{p,P}E_{p,S}$ term of Eq.\ \pref{finalEL}, since a helicity change corresponds to a sign change for one of the two pump components.  Beside the expected 4-fold symmetric contribution encoded in Eq.\ \pref{finalEL}, we detect a subleading 2-fold symmetric contribution, which may be related to an accidental reduced symmetry of the sample or to other effects not captured by our modelling of the whole detection process.
Finally, we point out that Eq.\ \pref{finalEL} provides a generalization for the 
%closely resembles the polarization and directional dependence of the 
specular nonlinear anisotropic polarization effect (SNAPE), already reported in cubic crystals for linearly polarized optical pump pulses in the parallel configuration \cite{popov1996pump, burgay95}.  

Despite the excellent agreement between the theoretical expression \pref{finalEL} and the measured angular dependence of the signal, marked deviations are observed for the temperature dependence of the signal. Fig.\ \ref{polariz}\textbf{c} shows the time-dependent traces in the cross-polarized configuration at fixed angle $\theta$ and two different temperatures. In particular, due to the large value of the band gap, it is hard to ascribe the  significant temperature variations observed experimentally to an analogous temperature dependence of the electronic $\chi^{(3)}$  tensor. Indeed, the insulating gap $E_g$ shows a modest temperature renormalization ($\sim 0.1$ eV) in this temperature range \cite{wuSTO2020}. Since far from the absorption edge the electronic $\chi^{(3)}$  is expected to scale as $\chi^{(3)}\sim 1/E_g^4$ \cite{stryland}, the gap variation could account at most for a 10$\%$ increase of the Kerr response between 150 and 300 K. In addition, besides the oscillating component at $\sim 2\O_p$, which is consistent with Eq.\ \pref{finalEL} following the squared pump field in time, a non-oscillating underlying background is found.
%shape of the signal is seen to change drastically when lowering the temperature. These observations suggest that, beside the bare electronic contribution described by Eq.\ \ref{finalEL}, which is approximately temperature independent at THz frequencies, other microscopic processes with the same polarization dependence should be taken into account in order to fully reproduce the experimental findings.

\section{Ionic Kerr effect}

Here we argue that, besides the EKE, the non-linear response behind the THz Kerr effect admits an additional ionic contribution, where the virtual electronic state is reached via (or decays to) an intermediate two-phonon excitation triggered by two photons of the THz pump pulse, see Fig.\ \ref{intro}\textbf{b}. The frequency of the TO$_1$ IR-active phonon mode softens from 3.2 to 1.8 THz when lowering the temperature from 380 to 150 K in the cubic phase of STO \cite{vogt95,yamada}. As a consequence, the IKE is maximized around room temperature, where the frequency spectrum of our THz pump pulse overlaps with the phonon mode, while it gets progressively suppressed by the phonon softening below the pump frequency under cooling. As detailed in Appendix \ref{appC}, two intermediate phonon processes are possible. In one case, denoted by $\Delta\Gamma^{ph,(a)}$, the primary excitation consists in two photons driving two phonons with a strength controlled by the phonon effective charge squared, and only after a virtual electronic state is reached. This process, sketched in Fig.\ \ref{intro}\textbf{b}, can be effectively accounted for by simply replacing each term $\bar E_{p, P/S}$ in Eq.\ (\ref{finalEL}) for the driving pump field with 
\be
\begin{aligned}
\lb{forceA}
\bar F^{(a)}_{P/S}(t_{pp}) \propto \int dt' D(t') \bar E_{p, P/S}(t_{pp}-t'),
\end{aligned}
\ee
where $D(t)\equiv \theta(t) e^{-\gamma_T t} \sin{(\omega_T t)}$ is the (single) phonon propagator, being $\gamma_T$ and $\omega_T$ the phonon broadening and frequency at temperature $T$. In the second process, denoted by $\Delta\Gamma^{ph,(b)}$, the intermediate electronic state itself delivers its energy to two IR phonons with opposite momenta, leading to an effective modulation of the $\chi^{(3)}$ tensor. This process can be described by replacing the 
%In order to obtain the $\Delta\Gamma^{ph,(b)}$ contribution associated with process ($b$), instead, since in this case the effective force is proportional to the squared pump field, one should replace the 
$\bar E_{p, P/S} \bar E_{p, P/S}$ product in Eq.\ (\ref{finalEL}) with
\be
\begin{aligned}
\lb{forceB}
\bar F^{(b)}_{P/S,P/S}(t_{pp}) \propto \int dt' P(t_{pp}{-}t') \bar E_{p, P/S}(t') \bar E_{p, P/S}(t'),
\end{aligned}
\ee
where $P(t) \equiv \theta(t) \coth \left( \frac{\hbar \omega_T}{2k_B T} \right) e^{-2\gamma_T t} \sin (2\omega_T t)$ is the two-phonon propagator \cite{caldarelli22, gabriele_nc21}. 
%In both cases, the resulting phonon-mediated contribution to the third-order polarization, leading to the IKE, is quadratic with respect to the pump pulse and linear with respect to the probe field, sharing the same general structure of Eq.\ \pref{polar}. 
We point out that $\Delta\Gamma^{ph,(a)}$ has no direct correspondence in conventional Raman experiments and can only be triggered by THz light pulses. Process $\Delta\Gamma^{ph,(b)}$, instead, can be interpreted as the low-frequency time-resolved counterpart of second-order Raman scattering \cite{siciliano}. The full measured quantity, accounting for both the EKE and IKE, then reads
\be
\Delta\Gamma = \Delta\Gamma^e+\Delta\Gamma^{ph,(a)}+ \Delta\Gamma^{ph,(b)}.
\lb{finalTOT}
\ee
For symmetry reasons, we expect the ionic part to preserve the same polarization dependence of the electronic contribution. However, the IKE has a  different temporal profile,  due to the time delay in the intermediate excitation of the IR-active phonons, and a marked $T$ dependence, allowing one to decouple the two effects.
\begin{figure}[t]
\centering
\includegraphics[width=8.5cm,clip=true]{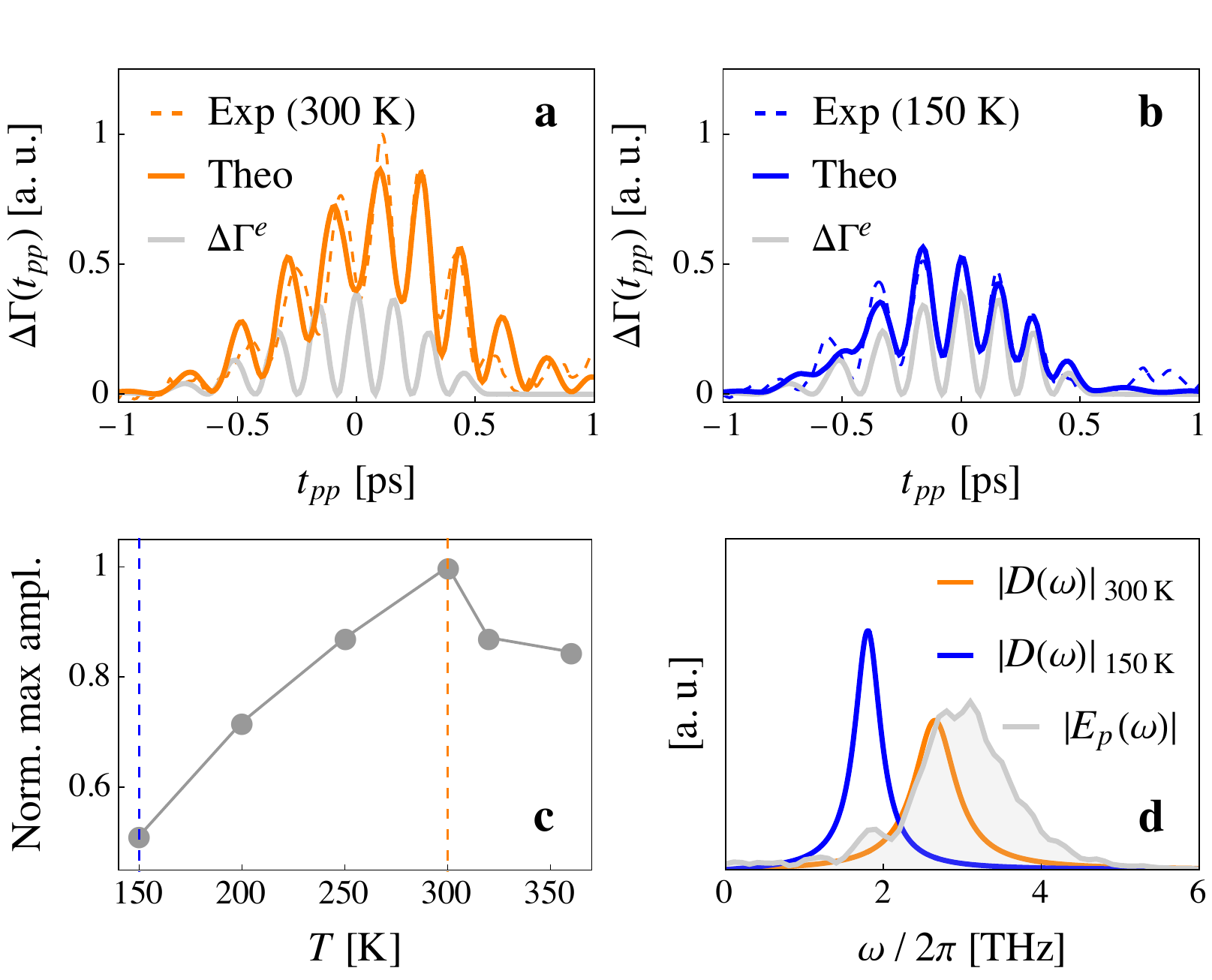}
\caption{Comparison between experimental data (dashed lines) and numerical simulations (\textbf{a}-\textbf{b}), in the absence (gray solid  line) and in the presence of phonon-mediated processes at 300 K (orange solid line) and 150 K (blue solid line).  \textbf{c} Full temperature evolution of the maximum in $\Delta\Gamma$ normalized to its $T=300$ K value. \textbf{d} Spectral content of the linear pump pulse (gray line) compared to the simulated Fourier transform of the IR-active phonon propagator at 300 K (orange line) and 150 K (blue line). Here we set $(\omega_T, \gamma_T)/2\pi \simeq (2.7, 0.3)$ THz at 300 K and $(\omega_T, \gamma_T)/2\pi \simeq (1.8, 0.2)$ THz at 150 K, in agreement with hyper-Raman measurements \cite{vogt95}.}  
\label{total}           
\end{figure}  
Experimental data and theoretical predictions at 150 and 300 K are compared in Fig.\ \ref{total}\textbf{a}-\textbf{b}. As expected, even if one assumes a temperature-independent magnitude of the electron-phonon coupling and of the internal pump field, the IKE is strongly reduced when lowering the temperature since the resonant condition between the soft-phonon mode and the pump field is lost, in good agreement with the experimental findings (see Fig.\ \ref{total}\textbf{c}-\textbf{d}). Above 300 K, instead, the phonon propagator overlaps with the pump spectrum, so that the relative reduction of the signal can be mainly ascribed to the screening of the THz pump pulse (see Appendix \ref{appD}). Nonetheless, both above and below 300 K the presence of a phonon contribution is further supported by a closer inspection of the spectral components of the signal, as detailed in Appendix \ref{appE}. Finally, it is worth noting that, while the velocity mismatch between the pump and probe pulses can influence in general the temporal shape of the Kerr signal at optical \cite{huber21ultrafast, maehrlein_PNAS} and THz \cite{frenzel23} frequencies in thick samples, propagation effects are expected to be negligible in our measurements due to the strong THz absorption close to the TO$_1$ phonon frequency \cite{nelson_sci} (see Appendix \ref{appD}). 

\section{Discussion and conclusions}
In the present manuscript we demonstrate a direct contribution of the soft IR active TO$_1$ phonon to the THz Kerr effect in insulating STO. The basic mechanism relies on the THz-driven resonant excitation of two IR-active phonons at $\o_1,\o_2$ frequencies, leading in general to a modulation of the refractive index at $\omega_1\pm\omega_2$. 
%While Raman-active phonons can be excited using eV pulses via a difference-frequency process, and as such they can contribute to the Kerr effect even with all-optical Kerr measurements, IR-active phonons can only be resonantly driven by THz light pulses, making the IKE a specific signatures of the THz Kerr effect \textcolor{blue}{(actually you can have process b with eV light!!!)}. 
In previously discussed non-linear phononic effects, the large lattice vibrations provide a way to drive other Raman-active or optically-silent lattice degrees of freedom anharmonically coupled to the IR-active resonant mode\cite{maehrlein_prb18,vonhoegen18probing,forst11nonlinear,subedi14theory,nelson_sci}. Here we additional demonstrate that they are responsible for a direct modulation of the conventional EKE. 
%The identification of the phonon contribution is made possible by a microscopic theoretical description of the whole detection process, via a quantitative description of the temporal and angular dependence of the THz Kerr effect measured via balanced detection. 
This general paradigm can be further extended to describe materials with different lattice symmetries and electronic properties, opening the avenue for the full exploitation of the IKE to investigate the electron-phonon coupling across various phase transitions. As an example, signatures compatible with the IKE have been recently observed in low-temperature KTaO$_3$ \cite{IKE_KTO}. This system, a parent perovskite material of STO, hosts a low-frequency TO$_1$ soft phonon mode that is supposed to play a crucial role for its magnetic \cite{balatsky_PRR, geilhufe} and superconducting \cite{norman_cm22} response. As such, further characterization of the IKE in this class of materials could provide insightful information on the materials properties in proximity to the ferroelectric and superconducting phase transition.

%\vspace{0.5cm}
%\begin{acknowledgments}
\section{Acknowledgments}
We are thankful to A.\ Geraldi and M.\ N.\ Gastiasoro for fruitful discussions. M.\ Basini and S.\ Bonetti acknowledge support from the Knut and Alice Wallenberg Foundation, grants No.\ 2017.0158 and 2019.0068. M.\ Udina and L.\ Benfatto acknowledge financial support by EU under MORE-TEM ERC-SYN (grant agreement No 951215) and by Sapienza University under project Ateneo 2022 No.\ RP1221816662A977. 
\\
%\end{acknowledgments}

%\textbf{Author contributions.} 
%L.B. and S.B. conceived the problem and supervised the work. M.B., M.P. and S.B. designed the experiment. M.B. and M.P. performed the experiments with the aid of V.U. M.U. and L.B. elaborated the theoretical model and carried out the numerical simulations with input from M.B. and M.P. All authors discussed the results. M.U., M.B. and L.B. wrote the manuscript with input from all authors.

\appendix
\section{Electro-optic characterization of THz pump pulses}
\label{appA}

The characterization of THz pump pulses has been performed via electro-optical sampling using a 50 $\mu$m-thick GaP crystal cut along the [110] crystallographic direction and placed at the sample position. Multiple THz reflections inside GaP can be observed starting from $t \sim 0.5$ ps in Fig.\ \ref{EOS}(a). Nonetheless, this has no influence on the measurements shown in the main text, since the large absorption at THz frequencies in STO suppresses multiple reflections below the experimental noise level. Therefore, we considered the cut trace (up to $t \simeq 0.5$ ps) to simulate $\Delta\Gamma$ in the $t_{pp}$ time-domain. For comparison, Fig.\ \ref{EOS}(b) shows the $\bar E_{p,S}$ and $\bar E_{p,P}$ orthogonal components of the circular pump pulse, obtained by placing a THz quarter-wave plate after the band-pass filter.  

\begin{figure}[h!]
\centering
\includegraphics[width=8.5cm, clip=true]{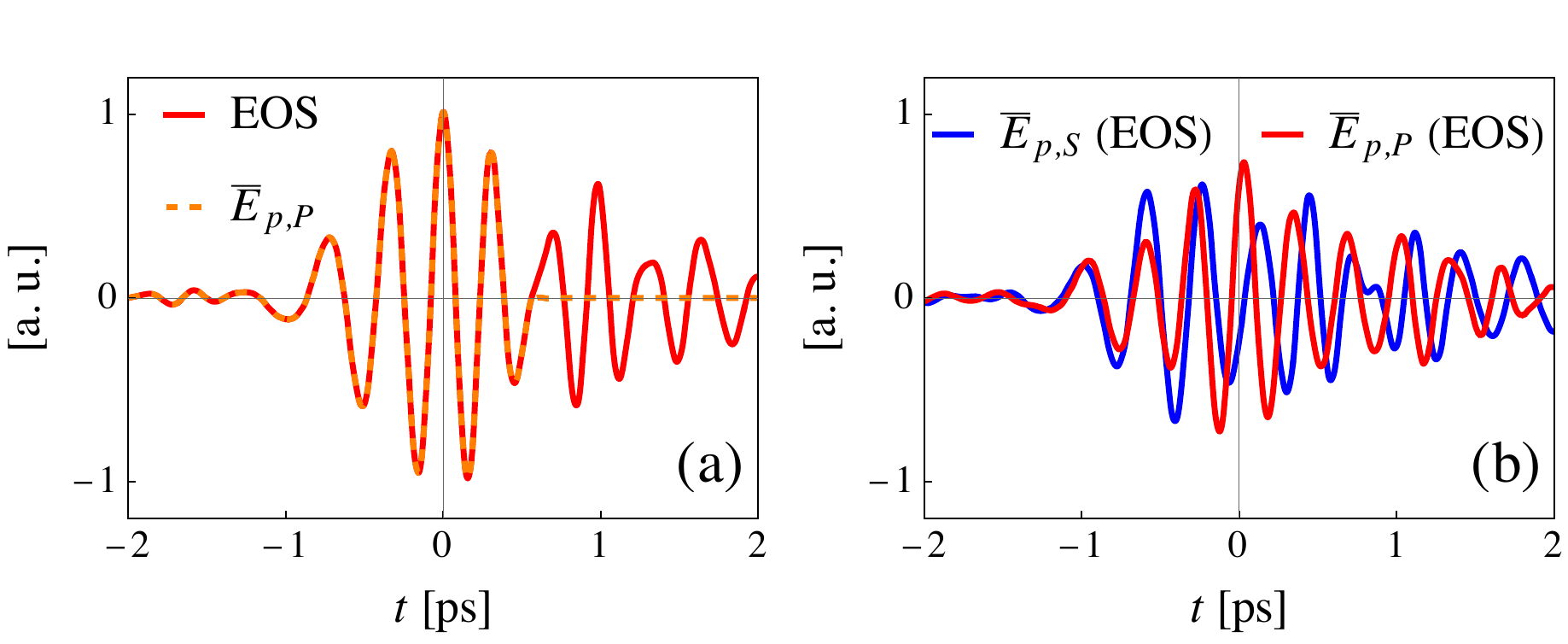}
\caption{(a) Electro-optical sampling (EOS) of the linearly polarized terahertz field normalized to its maximum value (red plain line) and the cut time trace used in the simulation (dashed orange line). The maximum field amplitude is 330 kV/cm  and the central frequency is $\O_p/2\pi =3$ THz. (b) Electro-optical sampling of the $\bar E_{p,S}$ (blue line) and $\bar E_{p,P}$ (red line) orthogonal components of the circularly polarized THz pulse, normalized as in (a). Here the maximum field amplitude is 230kV/cm, with central frequencies $\O_{p,P(S)}/2\pi \simeq 3.1(2.9)$ THz. }  
\label{EOS}
\end{figure}

\section{Pump-probe polarization misalignment}
\label{appB}

As mentioned in the main text, small deviations from the expected four-fold symmetric periodicity with zero average predicted by Eq.\ \pref{finalEL} are observed in the experimental data (see Fig.\ 3\textbf{a}). In particular, when linearly polarized pump pulses are applied, the signal presents an overall positive (in the cross-polarized configuration) or negative (in the parallel configuration) vertical shift. This slight deviation can be explained accounting for a non-zero polarization misalignment $\Delta_{\theta}$ between the pump and the probe pulses with respect to the $S$ direction. More specifically, the pump pulse $\mathbf{E}_p(t)$ components \pref{pump} can be rewritten as
\be
\lb{pumpDIS}
\mathbf{E}_p(t) = \begin{pmatrix}E_{p,S}(t) \sin(\theta+\Delta_{\theta})+E_{p,P}(t)\cos(\theta+\Delta_{\theta})\\E_{p,S}(t) \cos(\theta+\Delta_{\theta})-E_{p,P}(t)\sin(\theta+\Delta_{\theta}) \end{pmatrix},
\ee
while the probe field still forms an angle $\theta$ with respect to $y$ direction. Following the same steps detailed above, the measured quantity $\Delta\Gamma$ e.g.\ in the parallel linear configuration ($E_{p,P}=0$) becomes 
\be 
\begin{aligned}
\Delta\Gamma^e(t_{pp}^*,\theta, \Delta_{\theta}) \propto & \bar E^2_{p,S}(t_{pp}^*)\big[-\Delta\chi \sin(4\theta+2\Delta_{\theta}) +\\
+&\Delta\chi \sin(2\Delta_{\theta})\big],
\lb{snape2}
\end{aligned}
\ee
where the second term accounts for the additional isotropic contribution leading to the observed negative shift of the signal when $\Delta_{\theta}<0$. Such an explanation implies that the ratio between the overall constant shift and the maximum amplitude of the oscillating component should not depend on the spectral features of the pump pulse. This has been experimentally verified by slightly changing the alignment of incoming light in the parallel configuration, without filtering the broad-band THz pulse with the band-pass filter (see Fig.\ \ref{ang_dev}). The resulting vertical shift is consistent with the one observed in Fig.\ 3\textbf{a}, where instead narrow-band THz pulses were used.  
\begin{figure}[h!]
\centering
\includegraphics[width=5cm, clip=true]{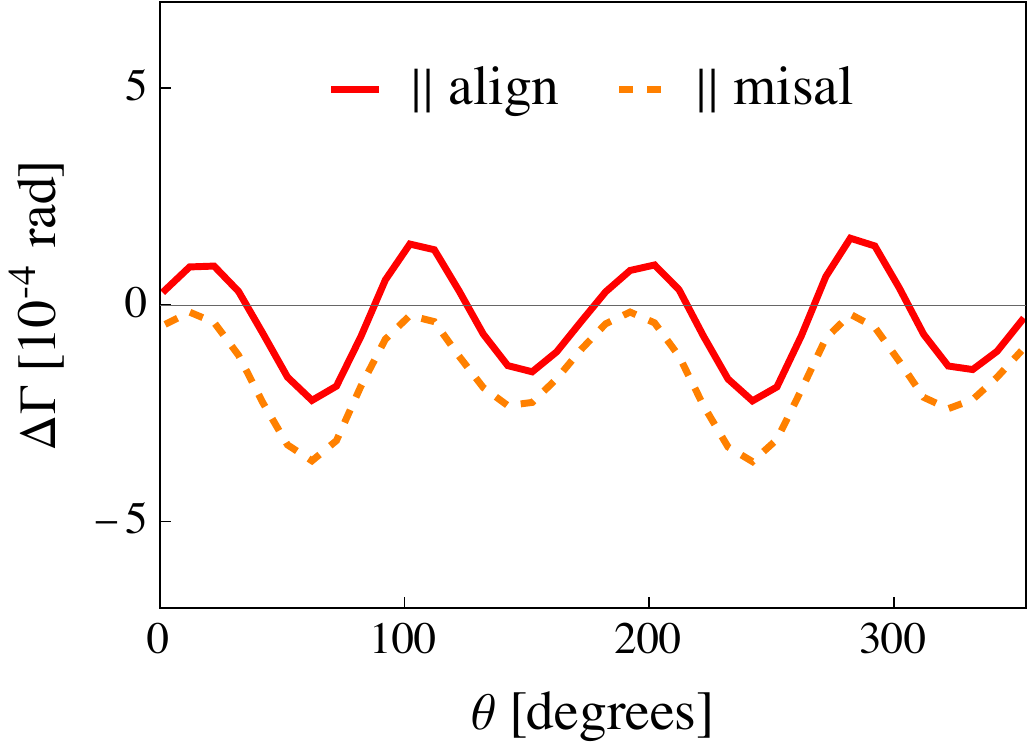}
\caption{Experimental data as a function of $\theta$ at fixed time-delay $t_{pp}$, obtained using broad-band THz pulses in the parallel ($\parallel$) configuration. When the pump pulse is aligned along the same direction of the probe (plain red line) the signal oscillates around zero. If one introduces a finite polarization misaligned with respect to the probe direction (here we set $\Delta_{\theta}\simeq -6^{\circ}$), instead, an overall vertical shift is observed (dashed orange line), in agreement with Eq.\ \pref{snape2}.}  
\label{ang_dev}          
\end{figure}  

\section{Explicit derivation of phonon-mediated contributions}
\label{appC}

Here we explicitly show how each contribution to Eq.\ \pref{finalTOT} can be associated with a third-order polarization sharing the same general structure of Eq.\ \pref{polar}, i.e.\ linear with respect to the probe field and quadratic with respect to the pump. From a diagrammatic perspective (see Fig.\ \ref{diagrams}), off-resonant interband electronic transitions leading to the EKE are associated with the bare electronic contribution $P^e$, coming from process (i), while the intermediate resonant excitation of the IR-active soft-phonon mode, responsible for the IKE, leads to the phonon-mediated contributions $P^{ph, (a)}$ and $P^{ph, (b)}$, associated with process (ii) and (iii) respectively. Each term can be obtained in a straightforward way by means of the effective-action approach, following the same steps detailed, e.g., in Ref.\ \cite{udina_prb19} for Raman-active phonon excitations in wide-band insulators. 
\begin{figure}[h!]
\centering
\includegraphics[width=8.5cm, clip=true]{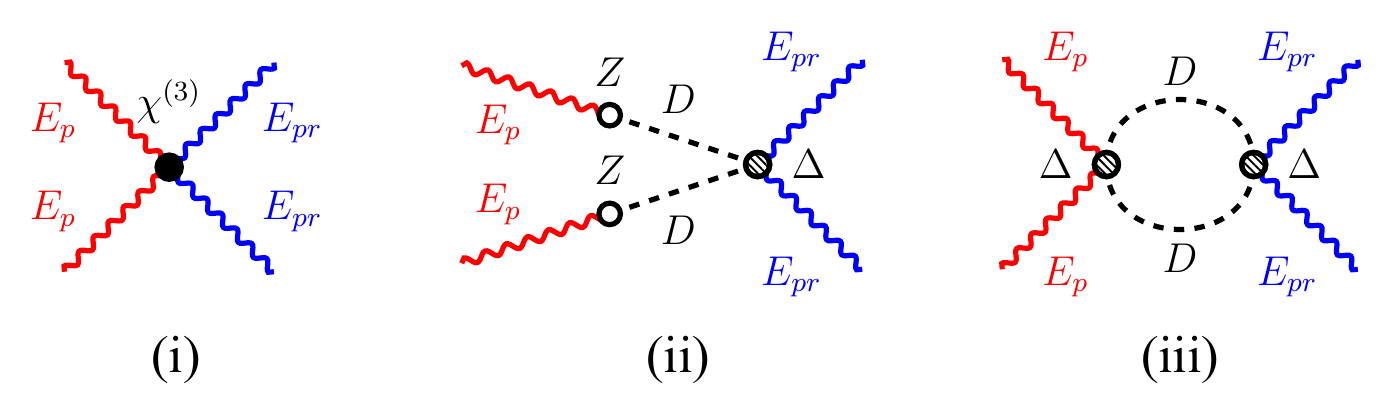}
\caption{Diagrammatic representation of the fourth-order contributions to the effective-action $S[E]$, associated with (i) off-resonant electronic transitions and (ii)-(iii) phonon-mediated resonant excitations. Red (blue) wavy lines represents the pump (probe) field, black dashed lines the phonon propagator. The filled dot represents the nonlinear electron susceptibility $\chi^{(3)}$, empty dots the phonon effective charge $Z$ and dashed dots the two-phonon effective coupling $\Delta$.}  
\label{diagrams}                                                   
\end{figure}  

In particular, the off-resonant electronic contribution $P^e$ can be related with the following action $S^e$ (see Fig.\ \ref{diagrams}(i)), at fourth-order with respect to the external electromagnetic field $E$:
\be
\begin{aligned}
\lb{s2}
S^{e}[E] \sim &
%\sum_{ijkl} \int \delta(t_1+t_3-t_5) \delta(t_2+t_3-t_5) \delta(t_4-t_5)  E_{p,l}(t_1)E_{p,k}(t_2)\chi_{ijkl}^{(3)}(t_3) E_{pr,j}(t_4)E_{pr,i}(t_5) dt_1dt_2dt_3dt_4dt_5\\
\sum_{ijkl}\int E_{pr,i}(t)E_{pr,j}(t)\chi_{ijkl}^{(3)}(t-t') \times\\
&\times E_{p,k}(t')E_{p,l}(t') dt dt' ,
\end{aligned}
\ee
which leads to the general Eq.\ \pref{polar} once the partial derivative with respect to the probe field has been performed, i.e.\ $P^e_i\equiv \partial S^e[E]/ \partial E_{pr,i}$. When off-resonant (electronic) transitions are considered $\chi_{ijkl}^{(3)}(t)\sim \chi_{ijkl}^{(3)}\delta(t)$. It is then useful to express the effective-action in the frequency domain, where the third-order susceptibility tensor is well approximated with the constant value $\chi_{ijkl}^{(3)}$. By performing the Fourier transform of each term, one then finds
\be
\begin{aligned}
\lb{s1}
S^{e}[E] \sim & \sum_{ijkl} \chi_{ijkl}^{(3)}\int E_{p,l}(\O_1)E_{p,k}(\O_2)\times \\
&\times E_{pr,j}(\O_3)E_{pr,i}(-\O_1-\O_2-\O_3)d\O_1d\O_2d\O_3,
\end{aligned}
\ee
where $\O_{1,2}(\O_3)$ are generic incoming THz(eV) frequencies taken from the pump(probe) pulse.

Focusing now on phonon-mediated processes, in the first case (Fig.\ \ref{diagrams}(ii)) two photons from the pump pulse linearly couple to the IR-active phonon mode before reaching the virtual electronic state, leading to the following contribution to the effective-action:
\be
\begin{aligned}
\lb{s3}
S^{ph, (a)}[E] \sim \sum_{ijkl} \int E_{p,l}(\O_1)Z_{l} D(\O_1) E_{p,k}(\O_2)Z_{k} D(\O_2) \times \\ 
\times \Delta_{lkij} E_{pr,j}(\O_3)E_{pr,i}(-\O_1-\O_2-\O_3)d\O_1d\O_2d\O_3.
\end{aligned}
\ee
Here $D(\O) \equiv \o_T/ [(\Omega+i\gamma_T)^2-\omega_T^2]$ is the phonon propagator in the frequency domain, $Z$ is related to the phonon effective charge and $\Delta$ gives the effective coupling between the two phonons and the (squared) probe field.
%, i.e.\ $\Delta_{\a\b ij} \sim \partial \chi^{(1)}_{ij}/\partial Q_{\alpha}\partial Q_{\beta}$, with $\chi^{(1)}$ the (linear) electric susceptibility and $Q$ the (IR-active) phonon displacement. 
While ab-initio estimates of $Z$ are routinely available in the literature, the estimate of the coupling function $\Delta$ requires state-of-art extensions of known DFT codes, that have been investigated only recently \cite{siciliano}. As a consequence, a detailed evaluation of the temperature and frequency dependence of $Z$ and $\Delta$ goes beyond the scope of this work. Nonetheless, we  notice that process (ii) should have in general the same polarization dependence of the bare electronic contribution (i), since $\delta \chi_{ijkl}^{(a)}\equiv Z_{l} Z_{k} \Delta_{lkij}$
%
%\be
%\lb{simpl}
%\delta \chi_{ijkl}^{(a)}\equiv \sum_{\a\b}Z_{l\a} Z_{k \b} \Delta_{\a\b ij}
%\ee
%
shares in principle the same tensor structure of $\chi_{ijkl}^{(3)}$. By then introducing the effective force $F_i^{ph,(a)}(\O) \equiv E_{p,i}(\O) D(\O)$, Eq.\ \pref{s3} can be rewritten as
\be
\begin{aligned}
\lb{s4}
S^{ph, (a)}[E] \sim & \sum_{ijkl}\delta \chi_{ijkl}^{(a)} \int F_l^{(a)}(\O_1) F_k^{(a)}(\O-\O_1) \times \\
&\times E_{pr,j}(\O_3)E_{pr,i}(-\O-\O_3)d\O_1d\O d\O_3,
\end{aligned}
\ee
with $\O=\O_1+\O_2$, and the resulting third-order polarization in the time domain reads
\be
\lb{s6}
P^{ph,(a)}_i(t,t_{pp}) \sim \sum_{jkl}\delta \chi_{ijkl}^{(a)} E_{pr,j}(t) \bar F_k^{(a)}(t+t_{pp}) \bar F_l^{(a)}(t+t_{pp}),
\ee
where $\bar F_i^{(a)}(t) = \int dt' \bar E_{p,i}(t-t')D(t')$, being $\bar E_{p,i}(t)$ centered at $t=0$.

In the second case, instead, the squared pump field first excites the (virtual) electronic state, which subsequently decays in two IR-active phonons (Fig.\ \ref{diagrams}(iii)). Therefore, the effective coupling $\Delta$ describes both the interaction with the pump and with the probe pulses, and the resulting fourth-order contribution to the effective-action scales as $\delta \chi^{(b)}_{ijkl} \equiv \Delta_{ijkl} \Delta_{ijkl}$.
%
%\be
%\delta \chi^{(b)}_{ijkl} \equiv \sum_{\a\b} \Delta_{kl\a\b} \Delta_{\a\b ji},
%\ee
%
In the frequency domain, one then finds
\be
\begin{aligned}
\lb{s8}
S^{ph, (b)}[E] \sim \sum_{ijkl}\delta \chi_{ijkl}^{(b)} \int  E_{p,l}(\O_1)E_{p,k}(\O_2)P(\O_1+\O_2)\times \\
 \times E_{pr,j}(\O_3)E_{pr,i}(-\O_1-\O_2-\O_3)d\O_1d\O_2 d\O_3,
\end{aligned}
\ee
with $P(\O)$ the effective two-phonon propagator \cite{gabriele_nc21}, i.e.
\be
\lb{s7}
P(\O) \sim \coth \left( \frac{\hbar \omega_T}{2k_B T} \right) \frac{\o_T}{(\Omega+i0^+)^2-4\omega_T^2},
\ee
where we neglect the phonon dispersion for simplicity. One then replaces $i0^+ \to i2\gamma_T$ to account for the finite phonon life-time.
Both the effective broadening $2\gamma_T$ and the $\coth$ term, accounting for the thermal distribution of the phonon population, can be explicitly obtained by deriving $P(\O)$ at finite temperature $T$ using Matsubara formalism \cite{caldarelli22}.  
Going back to Eq.\ \pref{s8}, the corresponding third-order polarization in the time-domain reads
\be
\begin{aligned}
\lb{P_phB}
P^{ph,(b)}_i(t,t_{pp}) \sim \sum_{jkl}\delta \chi_{ijkl}^{(b)} E_{pr,j}(t) \times \\
\times \int dt'  P(t+t_{pp}-t') \bar E_{p,k}(t') \bar E_{p,l}(t').
\end{aligned}
\ee
Starting from Eq.\ \pref{s6} and Eq.\ \pref{P_phB}, one then follows the same steps leading to Eq.\ \pref{finalEL} in the main text in order to derive $\Delta \Gamma^{ph,(a)}$ and $\Delta \Gamma^{ph,(b)}$ respectively. Notice that, while $\Delta\Gamma^e$ strictly follows the squared pump field in time, $\Delta \Gamma^{ph,(a)}$ follows the squared convolution between the pump and the phonon propagator (see Eq.\ \pref{s6}), while $\Delta \Gamma^{ph,(b)}$ follows the convolution between the two-phonon propagator and the squared pump field (see Eq.\ \pref{P_phB}). Due to the intermediate excitation of the phonon modes, each term shows a different phase factor in the time domain (see Fig.\ \ref{single}), in analogy with sum-frequency ionic Raman scattering \cite{maehrlein_prb18}. Consequently, their combination leads to interference effects responsible for the temperature-dependent background signal observed in the experiments, in addition to the oscillating component at $\sim 2\Omega_p$. 

\begin{figure}[h!]
\centering
\includegraphics[width=8.5cm, clip=true]{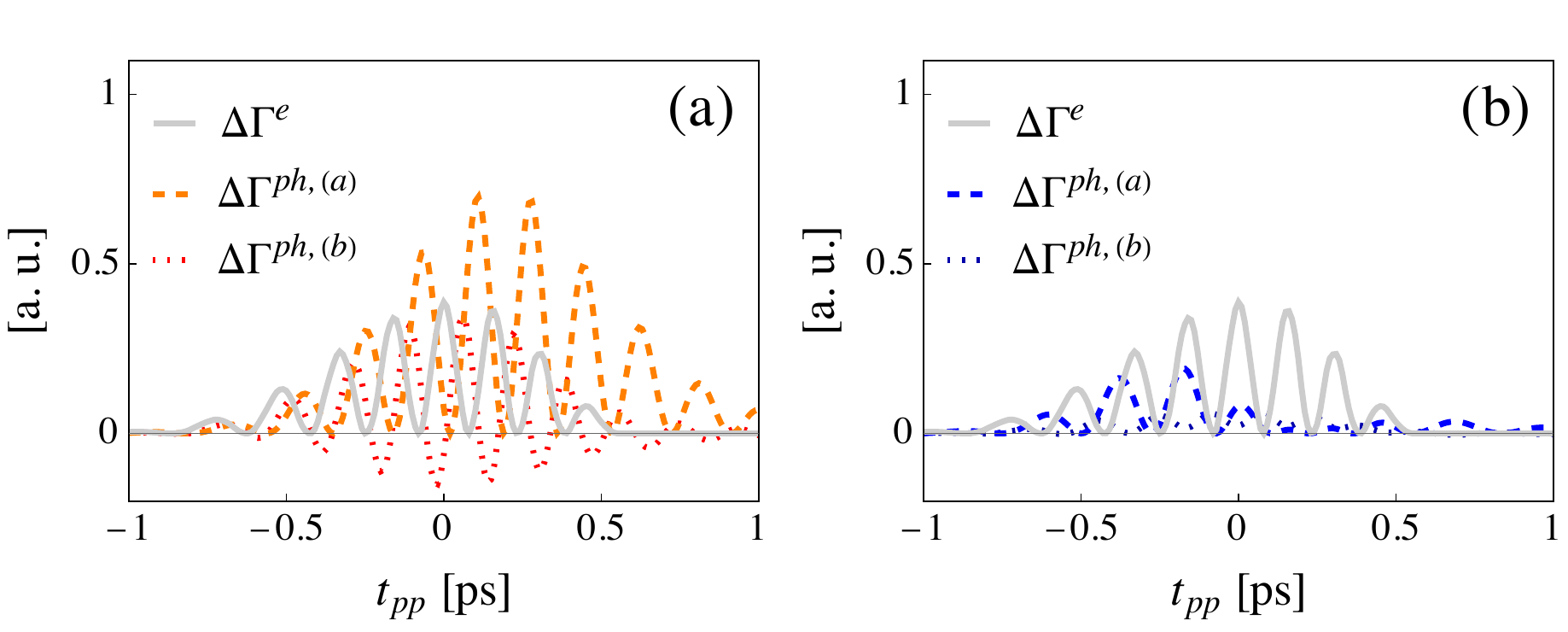}
\caption{Simulated single contributions to $\Delta\Gamma(t_{pp})$ at 300 K (panel (a)) and 150 K (panel (b)). While the off-resonant electronic contribution $\Delta\Gamma^e$ (plain gray line) is expected to be temperature independent, the phonon mediated processes $\Delta\Gamma^{ph,(a)}$ (dashed lines) and $\Delta\Gamma^{ph,(a)}$ (dotted lines) are strongly suppressed at 150 K, when the resonant condition between the phonon propagator and the pump spectrum is lost (see Fig.\ 4\textbf{d}). }
\label{single}                                   \end{figure}  

\section{Propagation effects and screening of the THz pump pulse}
\label{appD}

The enhanced value of the refractive index close to the soft-phonon frequency results in a sizable velocity mismatch between the THz pump and the optical probe pulses propagating inside the sample. In particular, at room temperature one finds $n+ik\sim2.5+i0$ at the probe frequency \cite{dejneka}, and $n+ik \sim 3.8+i6.4$ at the pump frequency \cite{vogt95, basini_DM}. This effect has been accounted for in Ref.\ \cite{sajadi15} by integrating the THz-field induced birefringence associated with the Kerr effect, i.e.\ $\Delta n(z,t)\propto E_p^2(z,t)$, along the propagating direction $z$ inside the sample:
\be
\label{eqPR1}
\Delta\Phi(t)\propto\int_0^L dz \Delta n(z,t+z/v_{pr} ),
\ee
where $L$ is the thickness of the sample and $v_{pr}$ the group velocity of the probe pulse. If the THz pump absorption inside the sample can be neglected, i.e.\ if $k\sim0$ and the pump field can be rewritten as $E_p (z,t)\to 2E_p (0,t-z/v_p )/(1+n)$, one finds that $\Delta\Phi(t)$ does not strictly follow the squared pump pulse $E_p^2(0,t)$ over time. Nonetheless, the soft-phonon mode leads to a strong THz absorption ($k\gg0$) in STO, which results in a very small penetration depth $l_{decay}$ of the THz pump field. In particular, at the pump frequency $\Omega_p/2\pi = 3$ THz one finds $l_{decay}\sim 2.5 \mu$m at 300 K and $l_{decay}\sim 3.6 \mu$m at 150 K \cite{basini_DM}. Since $l_{decay}\ll L=500 \mu$m, one can replace $E_p^2 (z,t)=\delta(z)E_p^2(z,t)$ in Eq.\ \pref{eqPR1}, i.e.\ the intensity of the incoming pump pulse is sizable only close to the sample surface ($z\simeq0$), the space integration gets simplified and $\Delta\Phi(t)\propto E_p^2 (0,t)$. In other words, the velocity mismatch between the pump and probe pulses is not expected to lead to significant timing modulations over the little length in which the THz field is sizable. This observation had been already pointed out in Ref.\ \cite{nelson_sci}, where analogous THz Kerr measurements had been performed on a 0.5 mm thick STO sample. 
To further confirm our expectations, measurements have been repeated by changing the central frequency of the THz pump pulse at fixed temperature $T=300$ K. When lowering the frequency, indeed, one expects the penetration depth to increase and, consequently, the effect of propagation to become more relevant. In particular, by modeling the dielectric function at THz frequencies as done e.g.\ in Ref.\ \cite{marsik16terahertz}, i.e.\
\be
\label{eqPR2}
\epsilon(\omega,T) \propto \epsilon_{\infty} + F^2 \frac{\omega_T^2}{\omega_T^2-\omega^2-2i\gamma_T\omega},
\ee
with $\omega_T$ and $\gamma_T$ the temperature dependent phonon frequency and phonon broadening, one finds that
\be
\label{eqPR3}
l_{decay}=\frac{1}{2\pi\omega k(\omega,T)}, 
\ee
where $k=\sqrt{(| \epsilon |-\epsilon')/2}$, is $\sim 10(100)$ times greater at $2(1)$ THz with respect to its 3 THz value (see Fig.\ \ref{propag}). The enhanced values of $l_{decay}$ result in pronounced time-variations of the measured signal with respect to the corresponding incoming squared pump pulses $E_p^2(t)$ at $\Omega_p/2\pi=2$ THz and $\Omega_p/2\pi=1$ THz. In particular, one finds a long-lasting exponential decay in the time trace which partly (or even fully) buries the Kerr response. These results are fully consistent with the experimental findings reported in Ref.\ \cite{frenzel23}, where similar THz Kerr measurements have been performed in different perovskite compounds. At the 3 THz pump frequency used in our measurements such spurious effects are actually minimized and this explains why the birefringence actually follows the $E_p^2$ time-dependence. 
\begin{figure*}[t]
\centering
\includegraphics[width=12cm, clip=true]{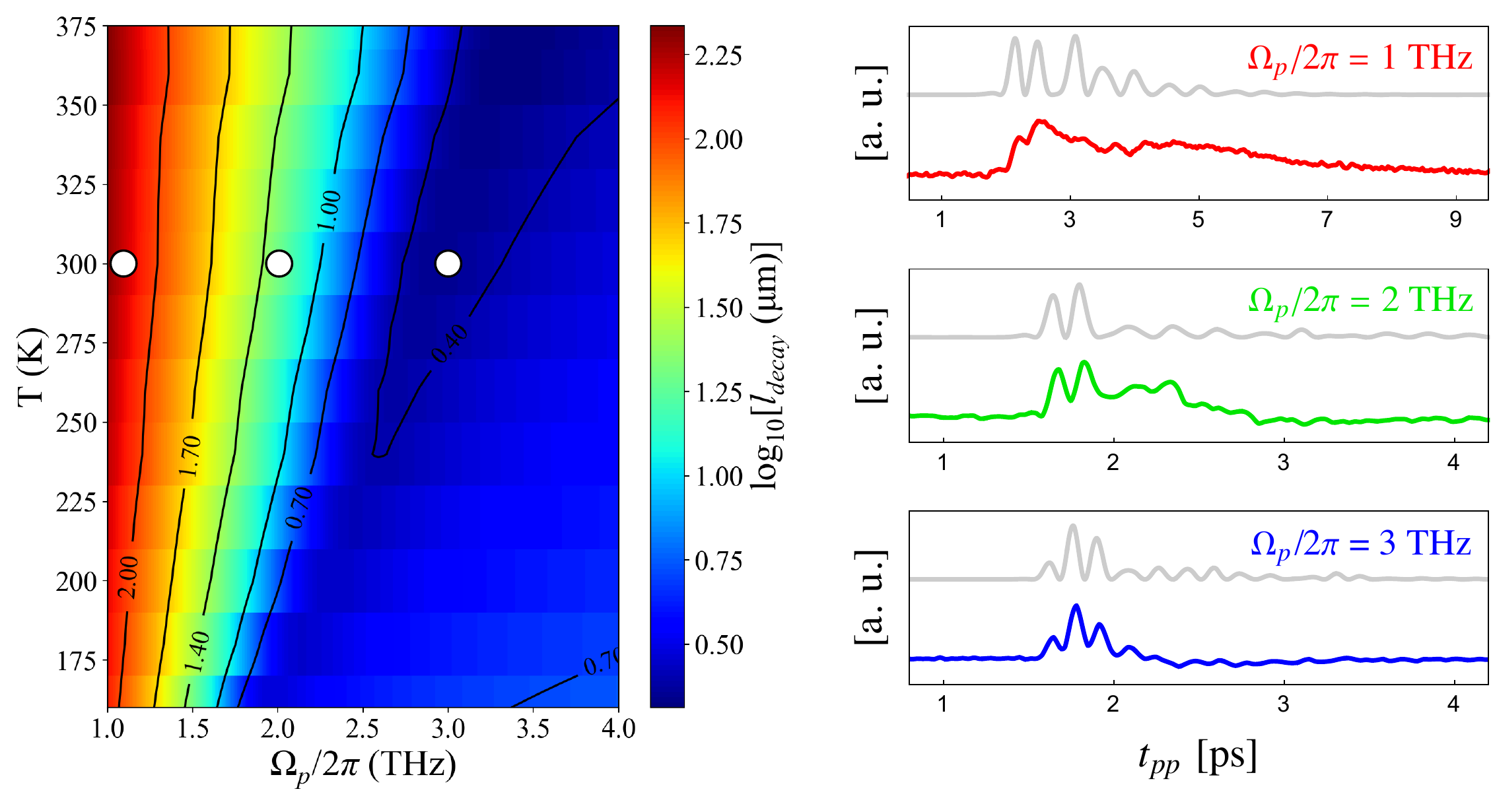}
\caption{Left panel: penetration depth $l_{decay}$ as a function of temperature and frequency as given by Eq.\ \pref{eqPR3}. The temperature-dependent phonon broadening and frequency are taken from Ref.\ \cite{basini_DM}. Right panel: measured differential intensity $\Delta\Gamma(t_{pp})$ at 300 K compared with the corresponding squared pump pulse $E_p^2(t_{pp})$, with different central frequencies $\Omega_p$ (gray lines). To obtain 1 THz pulses, we employed the organic crystal OH1 to generate broadband THz pulses within the frequency range of 0.5 THz to 3THz, and subsequently, we filtered them using a commercial (TYDEX) 1THz band-pass filter. To produce pulses centered at 2 and 3 THz, we employed the organic crystal DSTMS to generate broadband THz pulses within the frequency range of 0.5 THz to 6 THz and later filtered them using commercial (TYDEX) 2 THz and 3 THz bandpass filters. The corresponding penetration depths are marked by empty circles in the left panel. All curves are normalized to their maximum value.}  
\label{propag}                  
\end{figure*}  
\\

Beside propagation effects, one can further explore the consequences of a temperature-dependent screening of the THz pulse inside the sample by replacing $E_p(\omega)\to \tilde E_p (\omega,T)=2E_p(\omega)/[1+\sqrt{\epsilon(\omega,T)}]$. From Eq.\ \pref{eqPR2} one immediately notices that the spectral weight of the internal field $\tilde E_p(\omega,T)$ gets reduced when the pump pulse is resonant with the phonon-mode due to the enhanced THz absorption. Therefore, one would expect the amplitude of the Kerr response at 300 K to be smaller than at 150 K, in contrast with the experimental findings (see Fig.\ \ref{total}\textbf{c}). This observation further supports the presence of the ionic contribution to the Kerr response at 300 K, that is needed to overcome the detrimental effect of the internal screening on the electronic response. In contrast, for $T>300$ K the phonon spectrum overlaps with the pump spectrum, leading to a saturation of the phonon contribution and to an overall small decrease of the signal due to the linear screening (see Fig.\ \ref{screen}).
\begin{figure}[h!]
\centering
\includegraphics[width=8.5cm, clip=true]{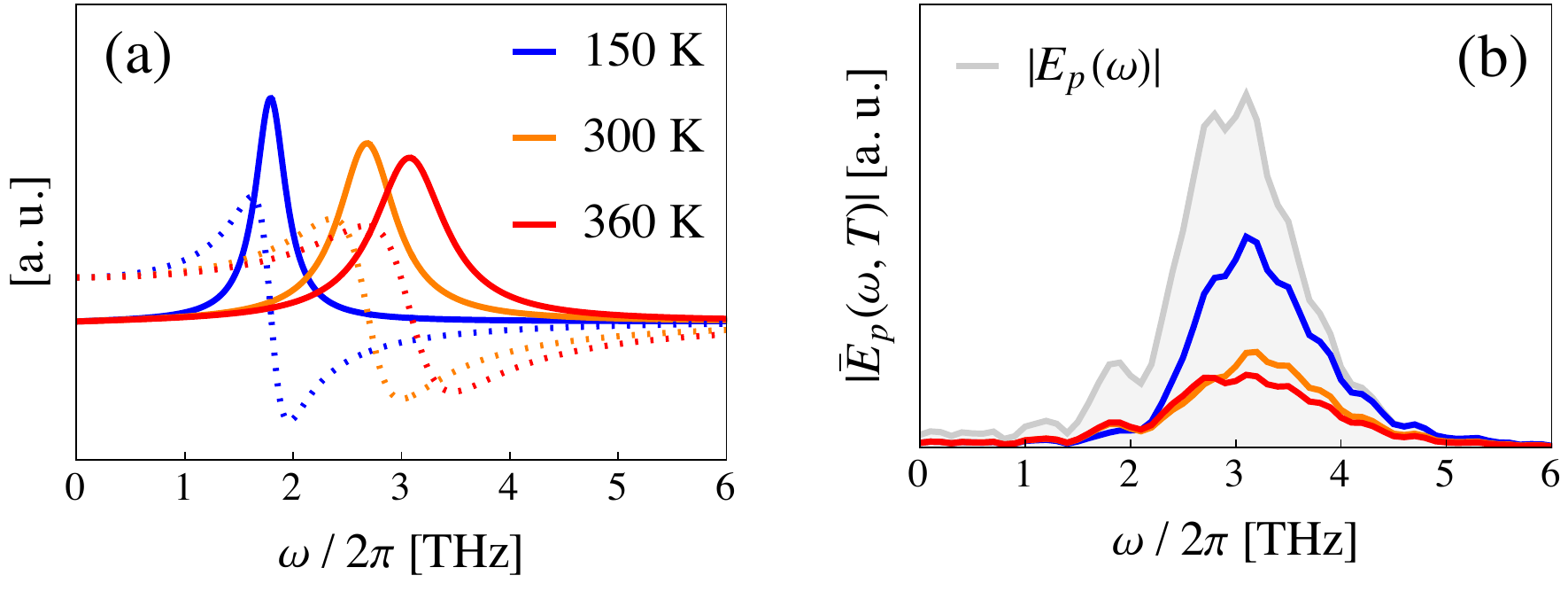}
\caption{(a) Real (dotted line) and imaginary (plain line) part of the dielectric function at THz frequencies, as given by Eq.\ \pref{eqPR2}, at 150 K (blue), 300 K (orange) and 360 K (red line). (b) Comparison between the absolute value of the external pump field (gray line) and the spectrum of the estimated internal pump field $|\bar E_p (\omega,T)|$ at 150 K (blue), 300 K (orange) and 360 K (red line).}
\label{screen}
\end{figure}

\section{Spectral content of the IKE}
\label{appE}

A clear indication of a phonon mediated contribution to the THz Kerr effect can be obtained by closer comparison between the spectral components of the Kerr response at different temperatures. Indeed, since the phonon peak softens when lowering the temperature, the IKE changes its spectral content. Below 300 K, the IKE survives only at frequencies where the phonon propagator overlaps with the pump spectrum. This leads to an enhancement of the signal at low frequencies as compared to room temperature measurements (Fig.\ \ref{spectr}(a)-(b)). On the contrary, at 360 K the phonon hardens above the pump frequency and the IKE contribution shifts at higher frequencies (Fig.\ \ref{spectr}(c)-(d)).

\begin{figure*}[t]
\centering
\includegraphics[width=11cm, clip=true]{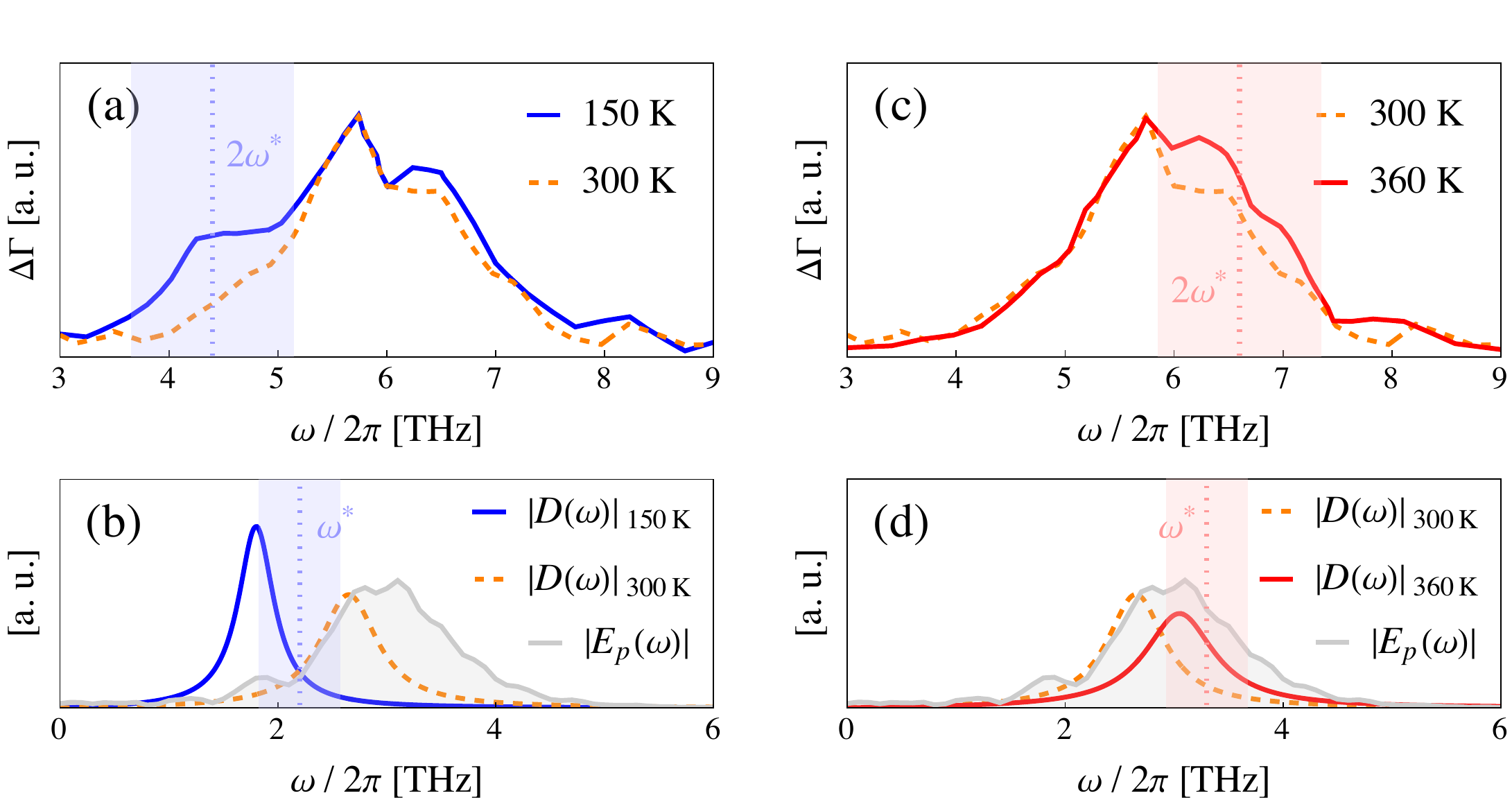}
\caption{Temperature evolution of the ionic contribution to the THz Kerr response in the frequency domain. (a) Normalized spectral content of the THz Kerr response at 300 K (orange dashed line) and 150 K (blue plain line). When lowering the temperature, the signal shows an enhancement around $2\omega^*\pm2\gamma^*$ (blue shaded region), with $\omega^*<\Omega_{p}$. (b) Pump spectrum (gray curve) compared with the simulated phonon propagator at 300 K (orange dashed curve) and 150 K (blue curve). At 150 K, the overlap between the phonon and the pump is maximized around $\omega^* \pm \gamma^*$ (blue shaded region), that corresponds to half of the frequency scale identified in panel (a). (c) Normalized spectral content of the THz Kerr response at 300 K (orange dashed line) and 360 K (red plain line). When increasing the temperature, the signal shows an enhancement around $2\omega^*\pm2\gamma^*$ (red shaded region), with $\omega^*>\Omega_p$. (d) Pump spectrum (gray curve) compared with the simulated phonon propagator at 300 K (orange dashed curve) and 360 K (red curve). At 360 K, the overlap between the phonon and the pump is maximized around $\omega^*\pm \gamma^*$ (red shaded region), corresponding in this case to the scale highlighted in panel (c).}
\label{spectr}
\end{figure*}

\bibliography{Literature3.bib}

\end{document}